\documentclass[%
 aip,
 amsmath,amssymb,
 reprint,%
]{revtex4-1}

\usepackage{graphicx}
\usepackage{dcolumn}
\usepackage{bm}

\usepackage[utf8]{inputenc}
\usepackage[T1]{fontenc}
\usepackage{mathptmx}
\usepackage{etoolbox}

\makeatletter
\def\@email#1#2{%
 \endgroup
 \patchcmd{\titleblock@produce}
  {\frontmatter@RRAPformat}
  {\frontmatter@RRAPformat{\produce@RRAP{*#1\href{mailto:#2}{#2}}}\frontmatter@RRAPformat}
  {}{}
}%
\makeatother

\usepackage{xfrac}
\usepackage{braket}
\usepackage{xcolor}
\usepackage{physics}
\usepackage{amsmath}
\usepackage{graphicx}

\newcommand{\StanfordAP}{Department of Applied Physics, Stanford University, Stanford, CA 94305, USA}
\newcommand{\StanfordPhysics}{
Department of Physics, Stanford University, Stanford, CA 94305, USA}
\newcommand{\PULSE}{
Stanford PULSE Institute, SLAC National Accelerator Laboratory, Menlo Park, CA 94025, USA}

\newcommand{\Davis}{Department of Chemistry, University of California, Davis, One Shields Avenue, Davis CA 95616}

\begin{document}

\preprint{AIP/123-QED}

\title[Fast array-based particle coincidence detection in a TimePix3-based velocity map imaging instrument]{Fast array-based particle coincidence detection in a TimePix3-based velocity map imaging instrument}

\author{Ian Gabalski}
 \affiliation{\PULSE}
 \affiliation{\StanfordAP}
 \affiliation{\Davis}
 \email{igabalski@ucdavis.edu}
\author{Eleanor Weckwerth}
 \affiliation{\PULSE}
 \affiliation{\StanfordPhysics}
 \email{eweck@stanford.edu}
\author{Chuan Cheng}
 \affiliation{\PULSE}
 \affiliation{\StanfordPhysics}
 \email{chengcc1@stanford.edu}
\author{Philip H. Bucksbaum}
 \affiliation{\PULSE}
 \affiliation{\StanfordAP}
 \affiliation{\StanfordPhysics}
 \email{phbuck@stanford.edu}

\date{\today}

\begin{abstract}
With the development of high repetition rate laser sources and advanced multi-particle correlation analyses such as covariance mapping, particle detection techniques such as velocity map imaging (VMI) are poised to offer unprecedented views into molecular phenomena. Taking full advantage of the high count rates in these experiments requires the development of detectors with sufficient spatial and temporal resolution that can process data in real time. The TimePix3 camera (TPX3CAM) is an event-based pixel detector capable of spatio-temporally localizing many simultaneous particle hits in an efficient manner. 
While the sparse nature of the data stream allows for compact representation of particle hits, it also presents algorithmic and computational challenges for clustering individual pixels into hits. 
Here we present the theory and application of a rapid data processing and centroiding algorithm for ion and electron hits collected in a VMI instrument. The array-based computations that comprise the algorithm take full advantage of the data sparsity of the TimePix3 data stream and localize particle hits on the microchannel plate (MCP) to better than a single pixel on the pixel detector. Centroiding can be parallelized on a commercially available graphics processing unit (GPU) for additional speed. Using these innovations, data processing occurs about 25 times faster than data acquisition, for a 1~kHz repetition rate instrument and tens of particles per shot. In addition to its speed, the TPX3CAM detector outperforms state-of-the-art delay line anode detectors at discriminating multiple simultaneous hits, enabling high-fidelity coincidence and covariance studies in the near future.

\end{abstract}

\maketitle

\section{\label{sec:intro}Introduction}

Velocity map imaging (VMI) spectroscopy~\cite{Eppink1997_Velocitya,Chandler1987_Twodimensional} is a powerful and widely used technique for resolving the momentum and angular distributions of charged particles produced in photo-induced processes. VMI detectors enable reconstruction of the three-dimensional momentum of each ion following a molecule's interaction with a sequence of laser pulses using position- and time-sensitive detection. Coincidence detection schemes, in which multiple ions produced in the same event are detected simultaneously, can further provide access to the complete kinematic information of a photo-induced process. When the molecular charge state is high enough, as is typically the case in strong-field- or X-ray-induced multiple ionization, the Coulomb explosion of the charged photofragments can be used to infer the instantaneous molecular geometry.~\cite{Burt2017_Coulombexplosion, Allum2018_Coulomb, Hasegawa2001_Coincidence, Bhattacharyya2022_StrongFieldInduceda, Boll2022_Xraya, Rudenko2017_Femtosecond, Unwin2023_Xray, Bisgaard2009_TimeResolved} VMI techniques can likewise be used to collect the photoelectrons produced by the ionizing pulse,~\cite{Gabalski2023_TimeResolved, Facciala2025_Unraveling, Horton2019_Excited} the energies and angular distributions of which provide information on the underlying molecular states and the properties of the photoionization process itself.~\cite{Gessner2006_Femtosecond, Hockett2011_Timeresolved, Gabalski2023_TimeResolved} Both electron and ion VMI techniques can be combined in a single experiment to produce a complete picture of the laser-induced molecular process.~\cite{Lehmann2012_Velocity, Zhao2017_Coincidenceb, Cheng2020_Momentumresolveda, Erk2018_CAMPFLASH, Jin2025_versatile} VMI spectroscopy is thus a powerful tool for investigating ionization, photoexcitation, and photodissociation dynamics.~\cite{Vredenborg2008_photoelectronphotoion, Ladda2025_Velocity} 

Many of the processes that VMI is well suited to study, such as enhanced strong-field ionization, isomerization, molecular elimination, and nonsequential double ionization,~\cite{Howard2023_Filmingb, Wells2013_Adaptivea, Li2025_Imaging, Weber2000_Correlated} occur with low probability and therefore require large data sets to obtain reliable statistics. In recent years, the repetition rates of available ultrafast laser systems have steadily increased, which has enabled the study of low-yield or highly complex processes that were previously inaccessible.~\cite{Tschentscher2023_Investigating, Leshchenko2020_Highpower, Heuermann2023_188, Meier2024_Compact, Luu2018_Generation, Kottig2017_Generation, Brahms2023_Efficient, Golibrzuch2025_ionimaging, Nitz2025_Multimass} In parallel, covariance mapping techniques permit operation in high count rate regimes by extracting particle correlations even when multiple competing channels are present.~\cite{Frasinski1989_Covariance, Frasinski2016_Covariance, Pickering2016_Communication, Allum2021_MultiParticle, Cheng2023_Multiparticle} Together, these advances have provided access to processes that occur only rarely and were previously beyond experimental reach by allowing substantially more data to be collected in a shorter period of time. With these new capabilities comes a growing need for detection techniques and data processing pipelines that can keep pace with increasing data collection rates while maintaining spatial and temporal resolution of particle hits.

The particle detection mechanism of a VMI instrument consists of a microchannel plate (MCP) paired with either a phosphor screen or delay line anode (DLA) detector.
The MCP amplifies the signal generated by each incident particle, converting it into an electron cascade that can be efficiently detected.~\cite{Wiza1979_Microchannel} 
DLA detectors provide precise spatial localization and timing information of amplified particle hits by measuring the timing of electrical signals from crossed, overlapping wire grids.\cite{Jagutzki2002_Multiplea, Lampton1987_Delaya, Jagutzki2002_broadapplication} They are capable of high throughput by performing much of the signal processing in fast analog electronics, making them well suited for true coincidence measurements at high repetition rates. However, DLA detectors are fundamentally limited in their ability to operate in the covariance regime, where high particle event rates lead to detector saturation and ambiguity in hit assignment. State-of-the-art DLA detectors can localize simultaneous hits that are separated by several millimeters on the MCP, but they fail to distinguish hits that fall within this range.~\cite{Jagutzki2002_Multiplea} In contrast, optical detection using a phosphor screen enables simultaneous localization of many particles with high spatial resolution, making it better suited for experiments with high count rates. This approach shifts much of the data processing burden to software and often requires specialized data treatment for full three-dimensional momentum reconstruction~\cite{Lee2014_Coincidence} or else selects a particular ion time-of-flight range via electrical gating of the MCP.~\cite{Townsend2003_Direct} While the superior multi-hit resolving capabilities of optical detectors has motivated their use in VMI instruments, they struggle to keep pace with the multi-kHz repetition rates of modern laser systems.

These considerations have motivated the development and adoption of pixelated optical detectors capable of recording individual particle hits at high rates. One such detector is the TPX3CAM, an event-based optical camera with a temporal resolution of 1.5~ns that reports events over threshold from individual pixels.~\cite{Frojdh2015_Timepix3} 
Unlike conventional frame-based cameras, the TPX3CAM records data only when the signal in a pixel reaches a threshold, producing a sparse data stream where only pixels associated with particle hits are recorded. This allows the TPX3CAM to be used at high count and repetition rates without producing overwhelming amounts of data, as has been demonstrated in some recent experiments.\cite{Golibrzuch2025_ionimaging,Nitz2025_Multimass,Long2017_Ionion}
However, making full use of this capability requires determining which signals on the TPX3CAM detector correspond to each individual particle hit. 
Particle hit identification, also referred to as centroiding, is an essential step to recover full kinematic information in VMI data analysis, but it becomes increasingly complex as the count rate increases. These same experiments that used TPX3CAM for its data sparsity either resorted to building image frames from the data stream for conventional centroiding,\cite{Golibrzuch2025_ionimaging} centroided the data offline due to slow processing speed,\cite{Long2017_Ionion} or else neglected to centroid the data entirely.\cite{Nitz2025_Multimass}

In this work, we introduce a fast centroiding algorithm for a VMI detector that leverages the sparse representation of TimePix3 data to efficiently localize particle impacts with sub-pixel precision while remaining compatible with high count rate operation. This algorithm can be extended to a graphics processing unit (GPU) to parallelize the computation, enabling data processing rates that keep pace with modern high repetition rate laser systems. We demonstrate the performance of the algorithm on electron spectroscopy data collected in gas-phase argon ionized with 400~nm pulses, showing the sharpening of above-threshold ionization~\cite{Corkum1989_Abovethreshold} and Freeman resonance~\cite{Freeman1987_Abovethreshold} features upon centroiding. The processing speed of the algorithm is comparable to that achievable with DLA detectors, and the ability of the TPX3CAM to localize simultaneous hits is shown to outperform that of the hexanode DLA detector.~\cite{Jagutzki2002_Multiplea}

\section{\label{sec:apparatus}Apparatus}

The centroiding algorithm described in the following sections is implemented for the data collected from a voltage-switching VMI instrument similar to the apparatus described previously.~\cite{Zhao2017_Coincidenceb,Lehmann2012_Velocity} Laser pulses originating from a 1~kHz repetition rate Coherent Legend Ti:sapphire laser system are used to ionize gas-phase atomic and molecular samples in an ultra-high vacuum (UHV) chamber (P=1$\times10^{-9}$~Torr). The electrons and ions produced in the interaction are accelerated by an electric field from a system of four VMI electrodes towards an amplification and detection system consisting of a integrated MCP/P47 phosphor screen assembly (Hamamatsu). The electrodes rapidly switch polarity following each pulse to capture both the negative electrons and the positive ions on the detector, which are well-separated by their Time of Flight (ToF).~\cite{Zhao2017_Coincidenceb} The front and back of the MCP are held at 0~V and +1900~V, respectively, while the phosphor screen is held at +4000~V. The back face of the phosphor screen is imaged onto the TPX3CAM detector area by a compound camera lens. The electron and ion count rates vary depending on the experiment and laser parameters used, but the instrument is commonly operated under conditions reaching up to 50 particles/shot. This count rate is compatible with the localization of individual particle hits in all three dimensions (X, Y, ToF).

\section{\label{sec:background}TimePix3 Data Stream}
\begin{figure*}[ht]
    \centering
    \includegraphics[width=\textwidth]{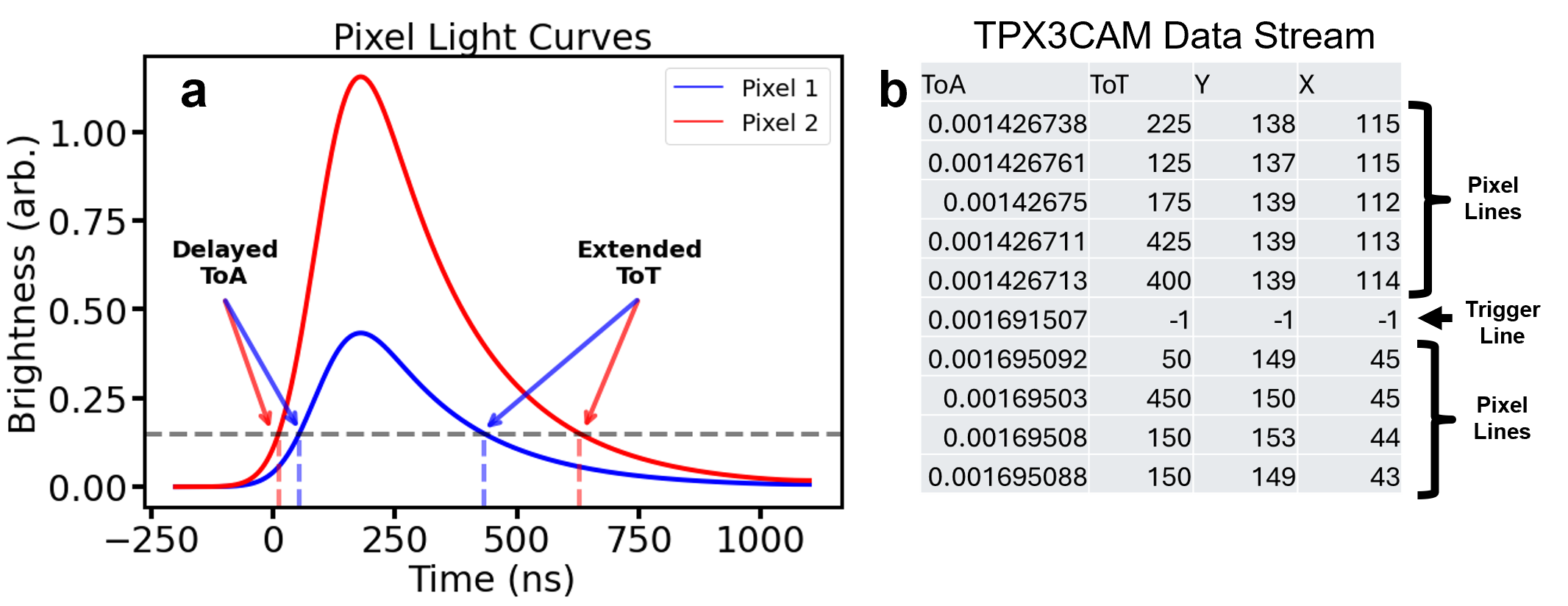}
    \caption[TimePix3 principles]{(a) Illustrated phosphor light curves (solid blue and red curves) along with TimePix3 threshold value (dashed gray line). TimePix3 pixels report their X and Y positions along with the time the signal first went over threshold (Time of Arrival, ToA), and the duration for which the pixel remained over threshold (Time over Threshold, ToT). (b) Example TPX3CAM data stream. Each line represents a pixel event or laser trigger.}
    \label{fig:light_curves}
\end{figure*}

TPX3CAM is a pixel detector with a straightforward silicon sensor substrate but remarkable and specialized read-out circuitry. Whereas typical CCD and CMOS cameras read out the entire array of pixels on a frame-by-frame basis, the pixels of the TimePix3 readout chip report data independently. When enough light enters a pixel to send it over its pre-set threshold, the pixel begins to collect data. Once the light intensity drops back below threshold, the pixel reports four separate pieces of data: its X and Y position on the detector array, the time at which the pixel initially went over threshold, and the duration for which the pixel remained over threshold. These four types of data are referred to in the following sections as X, Y, Time of Arrival (ToA), and Time over Threshold (ToT), respectively. A diagram of typical pixel light curves is shown in Figure~\ref{fig:light_curves}(a).

The TPX3CAM integrates the TimePix3 readout chip with a bump-bonded silicon detector that is anti-reflection coated for light in the range between 400~nm and 900~nm, which can effectively measure the 430~nm emission of the P47 phosphor.~\cite{Zhao2017_Coincidence} The sensor data stream is read out to a computer over a 10~GB/sec Ethernet connection. In addition to pixel data, the TPX3CAM has two ports for low-voltage TTL signals (3.3~V) that are processed by a time-to-digital converter (TDC). These ports can be used for laser triggers or other signals from lab equipment to determine the ToA of a pixel relative to such external signals. One of these TDC ports is used for the 1~kHz triggers from the laser system to determine the ToF of detected particles relative to the laser trigger for each shot. The data from the pixels and from the TDCs appear in the TPX3CAM data stream in line with each other as shown in Figure~\ref{fig:light_curves}(b), enabling convenient data processing within a single file. The data presented below are collected using SERVAL, a software developed by Amsterdam Scientific Instruments (ASI) to acquire data from TimePix3-based cameras like TPX3CAM. 
While the centroiding algorithm presented here can be applied more generally to any event-based detector, we focus in the remainder of this paper on the TPX3CAM data stream as a concrete example.

\section{\label{sec:processing}Preprocessing}

The TimePix3 data stream consisting of network packets from the pixel and TDC events are converted into text files readable by other processing programs using a slightly modified version of a C++ program provided by ASI. This preprocessing also presents an opportunity to apply some additional corrections to the data stream, as described in the following few sections.

\subsection{ToA Unwrapping}

The ToA data from the pixels and TDCs represents the elapsed time between the start of the data acquisition and the time the pixel or TDC went over threshold. This is represented digitally by the TimePix3 chip as an integer multiple of the fundamental temporal unit of the chip, equal to $25/4096$~ns, or about 6.1~ps. The pixels have a bit depth of 42, while the TDCs have a bit depth of 44. This means that after $25/4096\times2^{42}$~ns, or about 26.8~seconds, the pixel ToA will wrap back around to zero, while for the TDCs the ToA will wrap after $25/4096\times2^{44}$~ns or about 107.4~seconds. Since comparisons between ToAs of the laser trigger and ion hits must be made on each shot to determine ion ToF, this incongruous time wrapping must be corrected. An ``unwrapping'' of the ToA values is perfomed by looking for large negative shifts in the ToA of adjacent laser triggers and pixels separately. Each time the ToA of the next pixel or TDC is more than 20 seconds less than the previous one, the global counter of time wrappings is incremented by one and the appropriate integer multiple of fundamental time units is added to all subsequent data in the stream. 

This unwrapping must be done extremely carefully, and edge cases must be completely accounted for, as missed or incorrectly handled wrapping events will affect all subsequent data. One important edge case comes from the order in which pixel data are reported. A pixel only reports its data once its signal has dropped back below threshold, but it reports the ToA as the time at which it first went over threshold. In Figure~\ref{fig:light_curves}(a), the red light curve has an earlier ToA than the blue light curve, but because the blue curve is dimmer than the red it goes back below threshold first and reports earlier in the data stream. If the time wrapping occurs in between the two ToAs of the red and blue light curves, then the unwrapping due to the blue light curve will be incorrectly applied to the red light curve's ToA since it appears later in the data stream. To avoid this edge case, both large negative and large positive jumps are tracked in the ToA and the appropriate time unwrapping adjustment is applied in each case.

\subsection{Eliminating Background Counts}

For the apparatus described in Sec.~\ref{sec:apparatus}, the laser repetition rate was 1~kHz and typical ion ToF values did not exceed 10~$\mu$s. As a result, about 99\% of the acquisition window corresponds to ``dead time'' in which no laser-generated ions are expected to hit the detector. There are, however, sources of dark counts in the UHV chamber, such as the ion gauge used to track the chamber pressure during experiments. These dark counts occur at much lower rates than laser-generated ion counts when the laser is on, but due to the large fraction of dead time they can contribute a significant amount to the total generated data and drastically increase the size of the output file. This unwanted data is eliminated on initial processing by checking the ToF of each pixel with respect to the most recent laser trigger TDC value and only writing its data to the output file if its ToF does not exceed 100~$\mu$s.

\subsection{ToT-ToF Correlation Correction}

\begin{figure*}[ht]
    \centering
    \includegraphics[width=\textwidth]{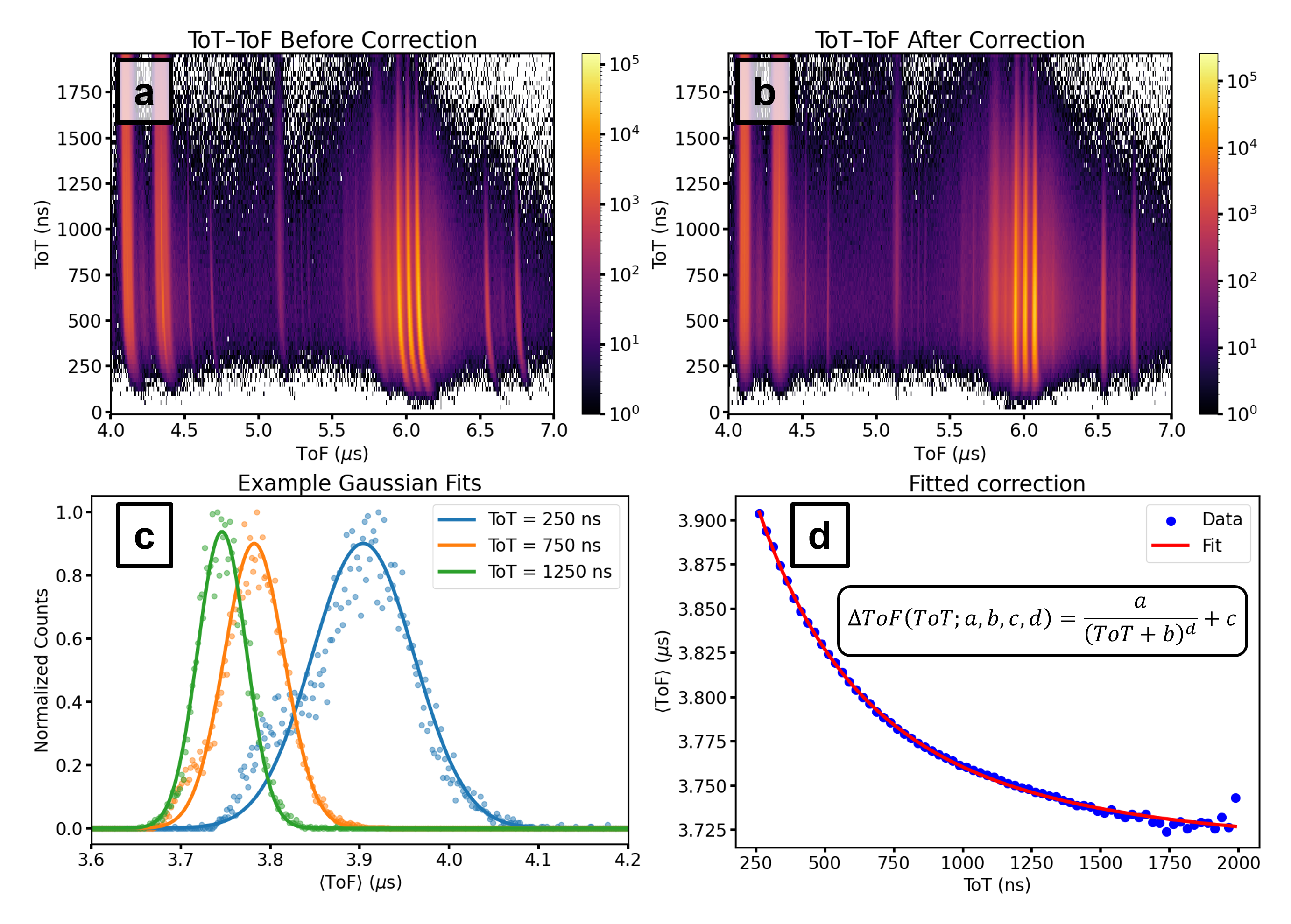}
    \caption[ToT-ToF Correlation Correction]{(a) Joint histogram of pixel ToT and ToF values for a typical data collection. The ToT-ToF correlation can be seen at the lowest ToT values, where the center of the distribution is shifted to larger ToF values. (b) Corrected ToT-ToF values after calibration. (c) Example Gaussian fits to the distribution of ToF values for individual ToT slices. The ToF center obtained from the fit moves earlier as the ToT value increases. (d) Plotting the ToF center $\langle \mathrm{ToF}\rangle$ obtained from the Gaussian fit vs. ToT gives the relationship between the two variables. This relationship is fit to the mathematical model shown in the figure and in Equation~\ref{eq:tottof}, the results of which are plotted in red.}
    \label{fig:tottof}
\end{figure*}

Figure~\ref{fig:light_curves}(a) illustrates how two light curves of identical intrinsic hit times but different brightnesses can have different ToA values as reported by the TimePix3 pixels. This induces a ToT-ToF correlation in which the dimmest pixels have ToF values that are shifted later in time than the rest, a phenomenon known as timewalk.~\cite{Turecek2016_USB,Tsigaridas2019_Timewalk} Figure~\ref{fig:tottof}(a) shows a joint histogram of the ToT and ToF values of pixels from a typical dataset. The correlation between ToT and ToF values can be seen at the lowest ToT values, where the center of each distribution is shifted to slightly larger ToF values. This correlation is corrected by fitting a model that maps the measured ToT and ToF values of each pixel to the true ToF value. This model is developed using two steps, illustrated in Figure~\ref{fig:tottof}(c-d). First, a particular ion peak is selected by taking a narrow ToF slice of the ToT-ToF histogram. Then, iterating over all ToT values in the slice, a Gaussian function is fit to the distribution in ToF to extract the center. A few examples of these Gaussian fits are shown in Figure~\ref{fig:tottof}(c), where it can be seen that the mean ToF value moves earlier for higher ToT values. The center of each Gaussian fit in ToF is plotted with the value of the ToT slice from which it was taken, as shown in Figure~\ref{fig:tottof}(d). A function~\cite{Turecek2016_USB} is fit to the result of the form
\begin{equation}
    \Delta ToF(ToT;a,b,c,d)=\frac{a}{(ToT+b)^d}+c
\label{eq:tottof}
\end{equation}
where $a$, $b$, $c$, and $d$ are free parameters. This mapping from measured ToF to actual ToF based on ToT is applied to every pixel in the data stream. The resulting ToT-ToF joint histogram after calibration is shown in Figure~\ref{fig:tottof}(b). In practice, the value of $d$ obtained from the fit is typically very close to 1, and thus a similar equation without this free parameter may be used instead. A single calibration can be used for all further data collection, but if the detection parameters such as VMI voltages or TimePix3 threshold values are changed, the parameters should be recalibrated.

\section{\label{sec:centroiding}Centroiding}

\begin{figure}[ht]
    \centering
    \includegraphics[width=0.5\textwidth]{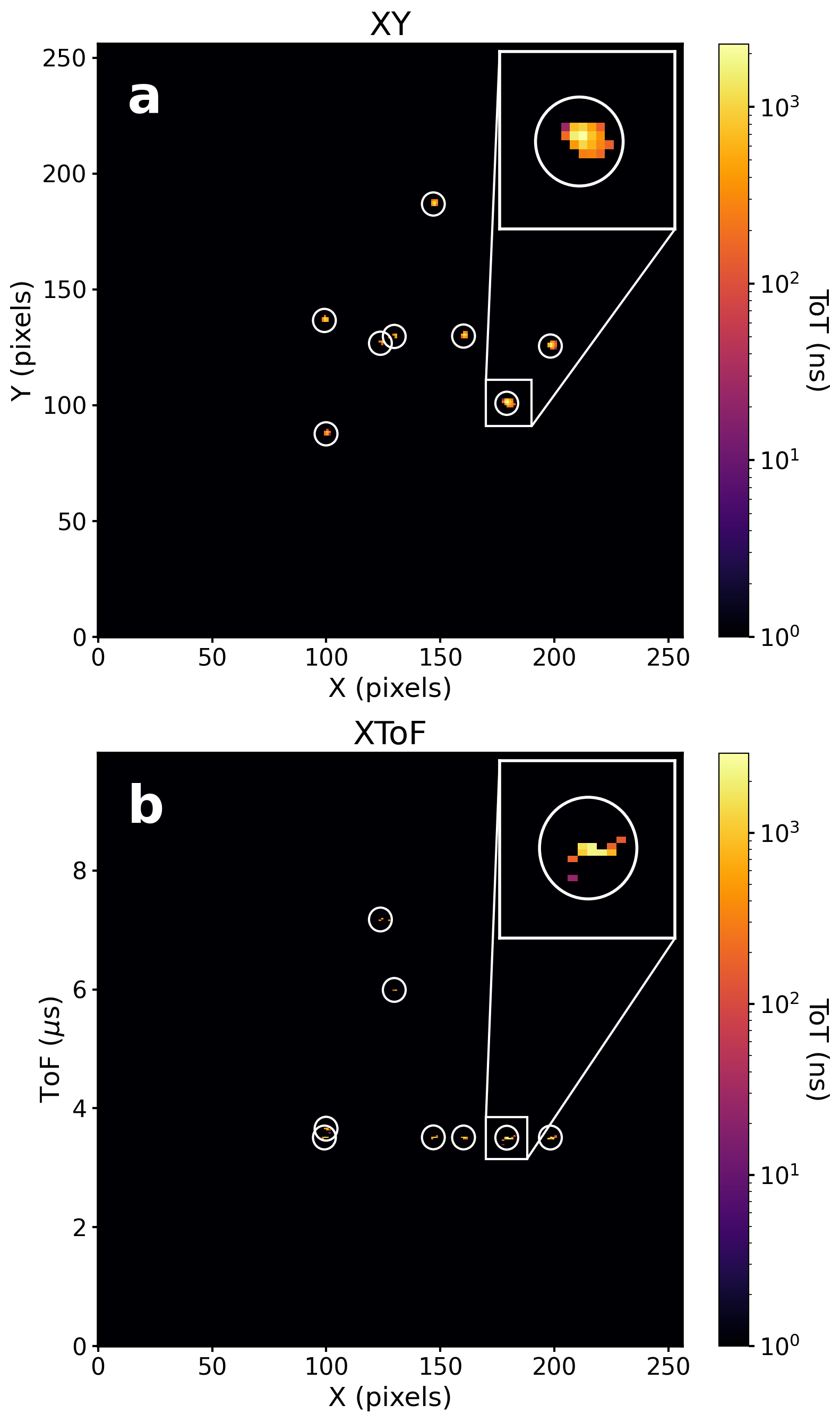}
    \caption[Example centroiding from a single laser shot]{Example of raw data and centroiding from a single laser shot. (a) X-Y histogram of all pixel data reported on a single shot, with brightness weighted by each pixel's ToT value. Inset: a zoomed particle hit to show distribution of pixel brightness. (b) X-ToF histogram of the same data in the previous panel. Each hit must be centroided in three dimensions, namely X, Y, and ToF. Inset: the same zoomed particle hit as shown in panel (a). Centers of the white circles represent the centroid locations.} 
    \label{fig:centroid_example}
\end{figure}

Due to the MCP amplification factor of approximately $10^{7}$ and the nature of the phosphor screen fluorescence, each particle hit can be extended over several pixels on the TimePix3 chip, depending on the particular detection parameters. In order to convert the raw data into hit-specific information, each extended spot must be centroided. Figure~\ref{fig:centroid_example}(a-b) shows histograms of the raw pixel data for a single laser shot in X-Y and X-ToF space, with Y-ToF not shown for brevity. Each particle hit must be centroided in three dimensions, namely X, Y, and ToF, in order to properly identify its species and momentum. The results of the centroiding process for this data are shown in Figure~\ref{fig:centroid_example}(c-d), with the white circles identifying the centroids of each particle hit. Identifying particle hits and obtaining their centroids from the raw pixel data stream  in a manner that fully leverages the sparsity of the TimePix3 data stream becomes a difficult problem to solve.

The primary challenge is that unlike a conventional camera that reports frames of pixel data in a regular grid each time, the TimePix3 camera reports a stream of pixel data that has no definite spatiotemporal relationship. This challenge also presents an opportunity for highly efficient data processing, as most of the pixels remain dark on each laser shot. The following scheme takes full advantage of the sparsity of the data stream in identifying individual hits and computing centroids of those hits, and because it consists of operations on arrays and matrices, it can be fully implemented on graphics processing unit (GPU) hardware to achieve drastic parallelization speedups.

The centroiding process is performed as a set of three steps:
\begin{enumerate}
    \item Neighborhoods: finding which pixels are ``neighbors'' with each other by computing the adjacency matrix
    \item Local Maxima: finding the brightest pixel in each neighborhood, which acts as the initial center of the centroiding computation and identifies how many particle hits are present
    \item Centroiding: computing the center-of-mass locations of each hit based on ToT values as a measure of brightness
\end{enumerate}
Each step is described in the following sections, including both the mathematical operation and its implementation as a computation on arrays and matrices.

\subsection{Neighborhoods}

Because the pixel data arrives in an essentially unsorted order, the first step is to find which pixel events are close enough to each other in X, Y, and ToF that they can be assumed to have originated from the same particle hit. We refer to this process of finding adjacent pixels as ``neighborhooding'' the data stream. Let $X_i$, $Y_i$, and $ToF_i$ be the $i$th X, Y, and ToF values for each pixel event in the data stream. The result of the neighborhooding process on the data stream is an adjacency matrix $A_{ij}$, where the $i,j$th entry in the matrix is either 1 or 0 and indicates whether pixel events $i$ and $j$ in the data stream can be considered to be in the same neighborhood as one another. 

Consider the neighborhooding problem in just one dimension, say X. As a concrete example, suppose the data stream $\{X_i\}$ consists of the X locations of each pixel event given by
\begin{equation}
    \{X_i\}=[20,21,90,92,18]
\end{equation}
There are two obvious neighborhoods, one centered at about 20 and another at about 91.
To compute the distance matrix $D_{ij}^X$, the repeated matrix $\{\{X_i\}\}$ is built and its transpose is subtracted.
\begin{equation}
\begin{split}
    D_{ij}^X&=\begin{pmatrix}
        20 & 21 & 90 & 92 & 18\\
        20 & 21 & 90 & 92 & 18\\
        20 & 21 & 90 & 92 & 18\\
        20 & 21 & 90 & 92 & 18\\
        20 & 21 & 90 & 92 & 18
    \end{pmatrix}-\begin{pmatrix}
        20 & 20 & 20 & 20 & 20\\
        21 & 21 & 21 & 21 & 21\\
        90 & 90 & 90 & 90 & 90\\
        92 & 92 & 92 & 92 & 92\\
        18 & 18 & 18 & 18 & 18
    \end{pmatrix}\\
    &=\begin{pmatrix}
        0 & 1 & 70 & 72 & -2\\
        -1 & 0 & 69 & 71 & -3\\
        -70 & -69 & 0 & 2 & -72\\
        -72 & -71 & -2 & 0 & -74\\
        2 & 3 & 72 & 74 & 0
    \end{pmatrix}
\end{split}
\end{equation}
A condition is then set on the distance being within a certain range, say $|D_{ij}^X|\leq5$.
\begin{equation}
    A_{ij}^X=\begin{pmatrix}
        1 & 1 & 0 & 0 & 1\\
        1 & 1 & 0 & 0 & 1\\
        0 & 0 & 1 & 1 & 0\\
        0 & 0 & 1 & 1 & 0\\
        1 & 1 & 0 & 0 & 1\\
    \end{pmatrix}
\label{eq:neighborhood}
\end{equation}
This neighborhooding operation is performed for each of X, Y, and ToF separately. The overall adjacency matrix $A_{ij}$ is obtained by computing the Hadamard product $\odot$, or element-wise product of matrices, of each of the three neighborhood matrices.
\begin{equation}
    A_{ij}=(A_{ij}^X)\odot(A_{ij}^Y)\odot(A_{ij}^{ToF})
\end{equation}

\subsection{Local Maxima}
The number of hits in a laser shot is counted and the approximate hit centers are obtained by finding the local maxima, or brightest pixel in each neighborhood. The measure of brightness used is the ToT value, because the brightest pixels also tend to remain over threshold for the longest. Consider again a data stream of five pixel events with adjacency matrix given by Equation~\ref{eq:neighborhood} and ToT data given by $\{ToT_i\}=[50,200,300,25,100]$. The ToT value of each pixel can be compared to just those of its neighbors by multiplying the repeated ToT array by the adjacency matrix and subtracting its transpose.
\begin{equation}
\begin{split}
    (ToT_{ij})\space \odot\space(A_{ij})&=\begin{pmatrix}
        50 & 200 & 300 & 25 & 100\\
        50 & 200 & 300 & 25 & 100\\
        50 & 200 & 300 & 25 & 100\\
        50 & 200 & 300 & 25 & 100\\
        50 & 200 & 300 & 25 & 100\\
    \end{pmatrix}\odot\begin{pmatrix}
        1 & 1 & 0 & 0 & 1\\
        1 & 1 & 0 & 0 & 1\\
        0 & 0 & 1 & 1 & 0\\
        0 & 0 & 1 & 1 & 0\\
        1 & 1 & 0 & 0 & 1\\
    \end{pmatrix}\\
    &=\begin{pmatrix}
        50 & 200 & 0 & 0 & 100\\
        50 & 200 & 0 & 0 & 100\\
        0 & 0 & 300 & 25 & 0\\
        0 & 0 & 300 & 25 & 0\\
        50 & 200 & 0 & 0 & 100\\
    \end{pmatrix}
\end{split}
\end{equation}
Subtracting the transpose of $(ToT_{ij})\space \odot\space(A_{ij})$ gives
\begin{widetext}
\begin{equation}
\begin{split}
    [(ToT_{ij})\space \odot\space(A_{ij})]-[(ToT_{ij})\space \odot \space(A_{ij})]^T&=\begin{pmatrix}
        50 & 200 & 0 & 0 & 100\\
        50 & 200 & 0 & 0 & 100\\
        0 & 0 & 300 & 25 & 0\\
        0 & 0 & 300 & 25 & 0\\
        50 & 200 & 0 & 0 & 100\\
    \end{pmatrix}-\begin{pmatrix}
        50 & 50 & 0 & 0 & 50\\
        200 & 200 & 0 & 0 & 200\\
        0 & 0 & 300 & 300 & 0\\
        0 & 0 & 25 & 25 & 0\\
        100 & 100 & 0 & 0 & 100\\
    \end{pmatrix}\\
    &=\begin{pmatrix}
        0 & 150 & 0 & 0 & 50\\
        -150 & 0 & 0 & 0 & -100\\
        0 & 0 & 0 & -275 & 0\\
        0 & 0 & 275 & 0 & 0\\
        -50 & 100 & 0 & 0 & 0
    \end{pmatrix}
\end{split}
\label{eq:local_maxima}
\end{equation}
\end{widetext}
The diagonal elements of this matrix are necessarily zero, as are any elements that correspond to a pixel that is not in the neighborhood of the row's pixel. If the row corresponds to the pixel that is the local maximum of the neighborhood, then the maximum value in that row will be zero, while if it is not the local maximum, then the maximum value in that row will be greater than zero. The local maxima in this example are clearly pixel 2 and pixel 3, with ToT values of 200 and 300 exceeding the other values in their neighborhood. Therefore, the local maxima in each neighborhood can be found by finding the rows of the matrix in Equation~\ref{eq:local_maxima} that have a maximum value equal to zero. The number of unique particle hits can also be determined by counting the number of rows for which the maximum value is zero. In a single array-based operation, both the number of particle hits and their approximate centers are idenfied. These values are the basis for the centroiding calculation.

An important and highly common edge case is when two or more brightest pixels in a neighborhood have exactly the same ToT value. This is common because ToT is quantized in integer multiples of 25~ns by the TimePix3 readout chip. Finding the local maxima in each neighborhood in the fashion described above will result in an over-counting of particle hits whenever this occurs. To solve this, the degeneracy of the ToT values is lifted by a very small amount that guarantees that the ToT values will all be unique. The ascending array $0.001\cdot \{i\}$ is added to the ToT values $\{ToT_i\}$, ensuring that when two pixels have the same ToT value initially, the one that appears later in the data stream will have the higher shifted ToT value. After local maxima have been found, the same ascending array is subtracted from the ToT values.

\subsection{Centroiding}
After identifying the approximate centers of each particle hit and the neighborhood of pixels that belong to that hit, the centroid of each particle hit can be computed. The term ``centroid'' is synonymous with the center of mass, which is computed as
\begin{equation}
    r_{\mathrm{COM}}=\frac{\sum_i m_ir_i}{\sum_im_i}
\end{equation}
where $m_i$ and $r_i$ represent the mass and position of each object in the system, respectively. Just as the center of mass represents the mean location of all of the mass in a system, the centroid of each particle hit represents the mean location of every fluorescence photon produced by that hit. The ToT value of each pixel event is again used as a proxy for its brightness.

The centroid about every pixel is computed and the centroids are selected for the pixels corresponding to the local maxima. Mathematically, the centroids are equal to
\begin{equation}
    X^\mathrm{cent}_i = \frac{\sum_j(X_j)(A_{ij})(ToT_j)}{\sum_j(ToT_j)(A_{ij})}
\end{equation}
and the values $X^\mathrm{cent}_{i=\mathrm{loc. max.}}$ are selected as the true centroids about the local maxima. The centroid calculation is identical for the X, Y, and ToF degrees of freedom.

\section{\label{sec:gpu}GPU Implementation}
Because each of the steps in the centroiding process can be described in terms of array operations, they can easily be extended to graphics processing unit (GPU) hardware that offers massive parallelization of computation. Even simple array operations such as sums across a particular axis can be sped up by orders of magnitude by implementing them on a GPU. In addition, data from multiple laser triggers can be processed in parallel with the use of batching, which extends the data along one additional axis that represents the trigger index before computing the centroids. The only limitation in speedup when utilizing GPU batch processing is the limit of memory on the GPU. We implemented the parallelized centroiding algorithm using an NVIDIA Titan RTX GPU. Native access to the GPU from Python was achieved using the PyTorch library.~\cite{Paszke2019_PyTorch} 

\begin{figure}[ht]
    \centering
    \includegraphics[width=0.5\textwidth]{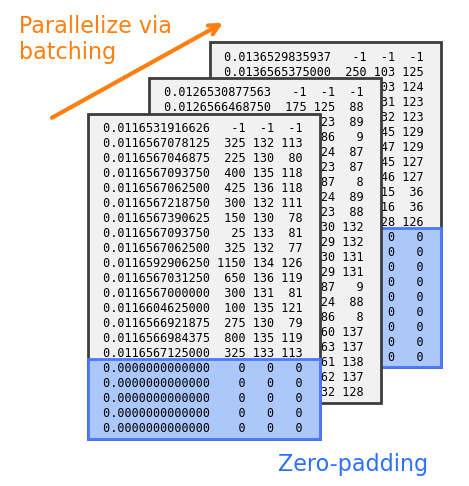}
    \caption[Batching of pixel data for centroiding]{Batching and zero-padding of multiple laser triggers of pixel data for centroiding. Several laser triggers worth of pixel data can be centroided simultaneously on the GPU by forming an array with one extra dimension along the laser trigger axis and applying all of the centroiding operations to the other axes. Zero-padding of the array data must be performed since each laser trigger contains a different number of pixel events.}
    \label{fig:batching}
\end{figure}

The primary challenge in achieving parallelization speedup via batch processing is that each laser trigger contains a different number of pixel events. Ragged arrays with different lengths along a particular axis are not permitted on GPU operations within PyTorch, so the ragged arrays must be zero-padded to form a rectangular array. A schematic of this zero-padding is shown in Figure~\ref{fig:batching}. When implementing computationally equivalent array operations in NumPy vs. PyTorch with GPU parallelization and batching, we achieve a speedup factor of $\sim5$x. In the data collection presented here on strong-field ionization of argon using a 1~kHz laser system and about 10~particles per shot, the centroiding performed on the GPU with batching ran about 25 times faster than data acquisition. 

The aspect of the computation that scales most unfavorably with count rate is the memory allocation of the neighborhooding operation. For a laser shot with $N$ pixels reporting data, the size of the adjacency matrix is $O(N^2)$. The quadratic scaling of memory required with (effectively) the per-shot count rate can rapidly outrun the available on-GPU memory. If high per-shot count rates are required, for instance if multi-body covariance measurements are of interest,\cite{Cheng2023_Multiparticle} then the number of laser shots over which the computation is batched can be reduced. On the other hand, if a choice can be made between running an experiment in low repetition rate, high per-shot count rate conditions vs. high repetition rate, low per-shot count rate conditions, such that the lab-time count rates are equivalent, the centroiding operation will be able to keep pace much more effectively with the low per-shot count rate, high repetition rate experiment.

\section{\label{sec:results}Results}
\subsection{Image Sharpening}

\begin{figure}[ht]
    \centering
    \includegraphics[width=0.38\textwidth]{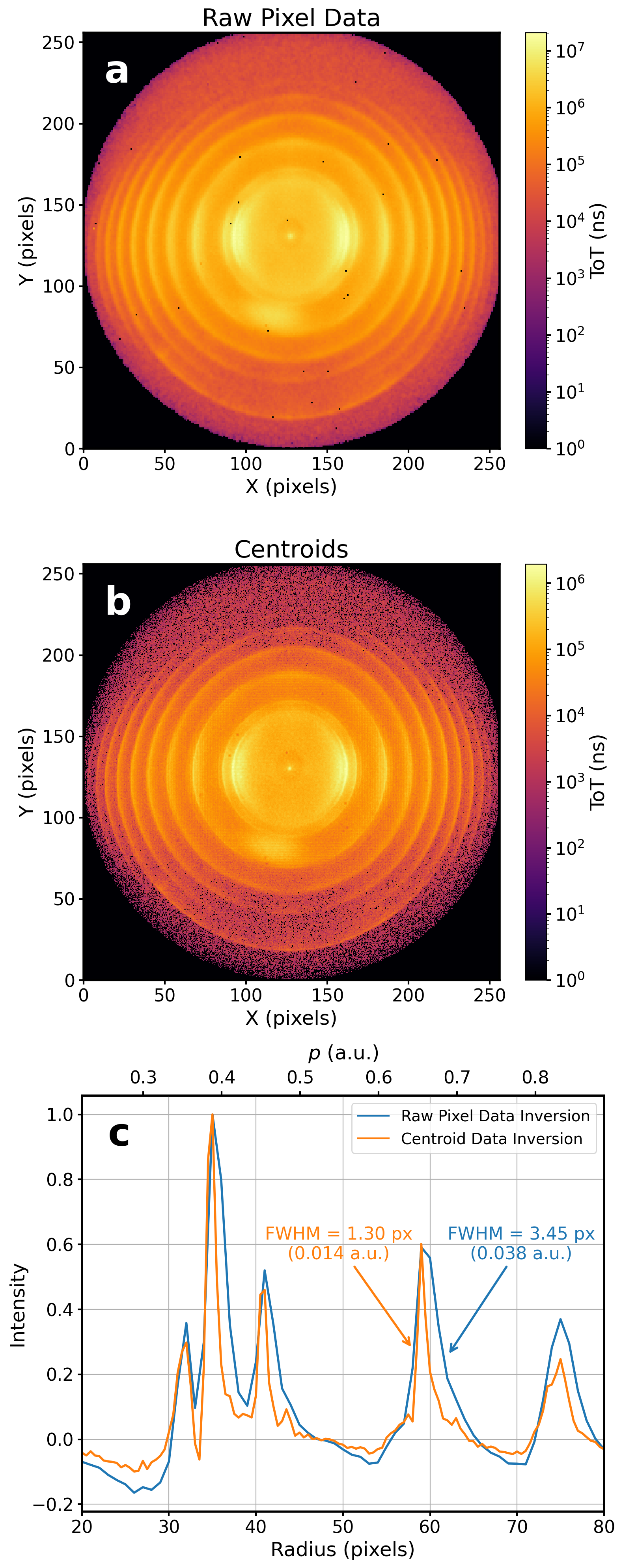}
    \caption[Comparison of raw pixel data and centroids]{Comparison of the X-Y histogram of electron hits from above-threshold ionization of argon at 400~nm wavelength for (a) raw pixel data, and (b) centroids with 0.5-pixel bin width. (c) Underlying photoelectron spectrum obtained via an inverse Abel transform performed on the raw pixel data and the centroided data. The various features in the photoelectron spectrum are sharpened by the centroiding.}
    \label{fig:centroid_raw_comparison}
\end{figure}

Centroiding provides significant sharpening of the spatial features of the data. Figure~\ref{fig:centroid_raw_comparison} shows a comparison of the X-Y histogram of electron hits generated via above-threshold ionization (ATI) of argon at 400~nm wavelength for the raw pixel data and for the centroided data. The raw pixel data are necessarily histogrammed in single-pixel bins, but the centroid data are histogrammed with 0.5-pixel bin width and show sub-pixel structure. While the ATI rings spaced by integer numbers of the photon energy are clear in both, the sub-structure of the electron momentum distribution, including both the azimuthal dependence and the emergence of sharp momentum features due to Freeman resonances,~\cite{Freeman1987_Abovethreshold} emerge from underneath the blurring effect of the multi-pixel fluorescence spot of the phosphor screen. Dead pixels present on the detector are effectively removed during the centroiding process as each particle hit spans multiple pixels, which can be seen by comparing the first two panels of Figure~\ref{fig:centroid_raw_comparison}. Figure~\ref{fig:centroid_raw_comparison}(c) shows the underlying photoelectron spectrum for the raw pixel data and centroid data computed via an inverse Abel transform based on polar onion peeling.\cite{Roberts2009_realtimea, Zhao2002_Deconvolvinga, Bordas1996_Photoelectronb} The image-sharpening effect of the centroiding is even more apparent, as both the first few ATI peaks above a radius of 50~pixels and the low-momentum Freeman resonance features below a radius of 45~pixels are much narrower. Low-intensity sub-structure that was blurred in the pixel data also becomes apparent upon centroiding. The width of the peak near 60 pixels from the center narrows by a factor of 2.6x after centroiding.

\subsection{Multi-hit Spatial Tolerance}

\begin{figure}[ht]
    \centering
    \includegraphics[width=0.5\textwidth]{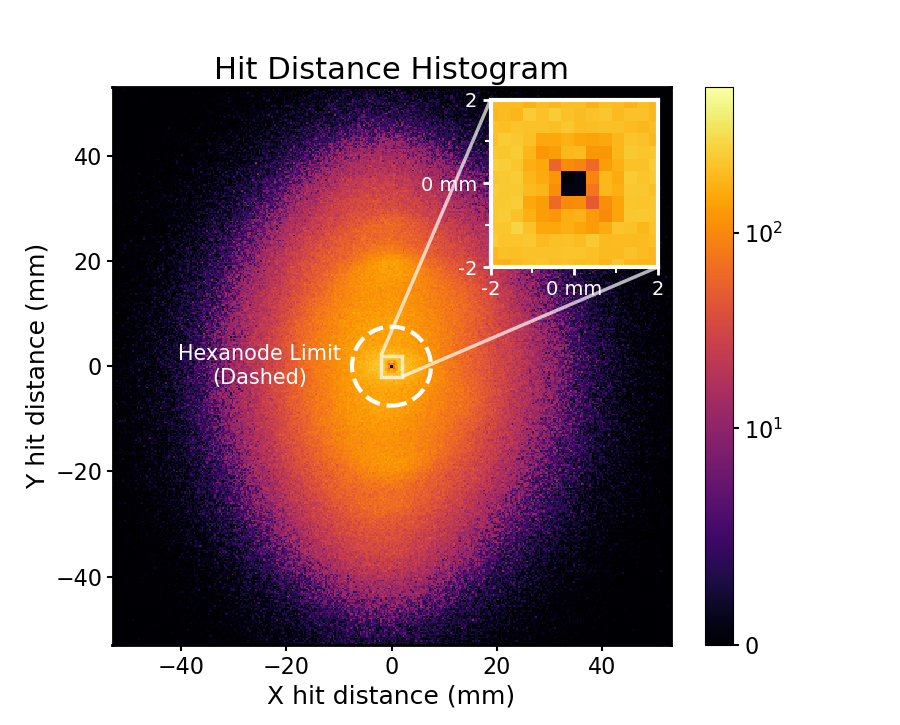}
    \caption[Histogram of hit distances]{Histogram of pair-wise centroid distances for all electrons from each laser shot in the argon ATI dataset. The minimum hit distance for the hexanode from Ref.~\cite{Jagutzki2002_Multiplea} is shown in the white dashed circle. (Inset) Zoomed histogram in the area close to zero distance between hits. The centroiding algorithm is capable of distinguishing simultaneous hits separated by anything greater than about 1~mm.}
    \label{fig:hit_distance_hist}
\end{figure}

The centroiding of individual particle hits is essential if multi-particle correlation methods such as coincidence and covariance are to be used. One of the primary drawbacks of DLA detectors is their inability to distinguish simultaneous hits that occur within some spatial separation on the detector due to the ambiguity of hit timing on each delay line. For the hexanode, a state-of-the-art DLA detector for multi-particle coincidence and covariance, the quoted minimum spatial distance required to distinguish simultaneous hits is around 7.5~mm.\cite{Jagutzki2002_Multiplea} Using the centroiding algorithm described here, the TPX3CAM is capable of resolving simultaneous electron hits from the argon ATI dataset that occur within a few mm of one another. Figure~\ref{fig:hit_distance_hist} shows a histogram of pair-wise separations between centroids from the same laser shot,  taken over about 500,000 laser shots from the 400~nm argon ATI dataset. For each laser shot, the photoelectrons all arrive essentially simultaneously in the VMI apparatus, and any pair of hits separated by less than about 7.5~mm would pose challenges for the hexanode DLA detector.~\cite{Jagutzki2002_Multiplea} Using the centroiding algorithm presented here, the TPX3CAM imaging the P47 phosphor screen is capable of distinguising pairs of simultaneous hits down to about 1~mm of separation on the detector surface. It should also be noted that this is not necessarily a fundamental limitation, as the optical imaging conditions of the TPX3CAM focusing lens as well as the voltage conditions of the MCP and phosphor can be modified to achieve even better multi-hit resolution. This drastic out-performance in coincidence hit resolving power, while not sacrificing on processing speed, motivates the use of the detection scheme and processing algorithm described here for high-count rate experiments in which coincidences of simultaneous hits are of interest.

\section{Conclusions}
We have described and demonstrated a fast array-based centroiding algorithm for use with a TPX3CAM detector on a VMI instrument that takes full advantage of the camera data stream's sparsity, processing data in real time with a 1~kHz laser repetition rate and up to tens of particles per shot. The algorithm is parallelized for rapid processing on a GPU and can be easily extended to higher laser repetition rates, keeping pace with the ever-advancing capabilities of modern laser sources. Particle hits are localized to better than a single pixel on the TPX3CAM sensor array, sharpening the detected spatial distributions of strong-field-ionized photoelectrons from argon gas. The detector is able to spatially distinguish multiple simultaneous hits that occur much closer together on the detector area than a hexanode DLA detector, the primary competitor to pixel detectors in terms of processing speed. The simultaneous performance of our particle coincidence detection in terms of processing speed for high repetition rate laser systems, spatial resolution, and compatibility with high coincidence count rates on a single laser shot, make the TPX3CAM-based detector an ideal candidate for modern VMI and other particle detection instruments.

\begin{acknowledgments}
    This material was based upon work supported by the U.S. National Science Foundation under Grant Number 2309238. Some processing code initially provided by Amsterdam Scientific Instruments, Inc. (ASI) was subsequently modified, as described in the text. We gratefully acknowledge A.M. Ghrist, H. Ma, A.J. Howard, L. Chaleunrath-Pham, E. Wells, S.A. Mohideen, and M. Britton for helpful discussions.
\end{acknowledgments}

\section*{Data Availability Statement}

The data that support the findings of this study are available from the corresponding authors upon reasonable request. The code implementation of the data processing algorithm described here is available on a Github repository: \url{https://github.com/Bucksbaum-Lab/TimepixCentroid}.

%

\end{document}